# The Planck High Frequency Instrument, a 3rd generation CMB experiment, and a full sky submillimeter survey

J.M. Lamarre[1], J.L. Puget[2], F. Bouchet[3], P.A.R. Ade[4], A. Benoit[5], J.P. Bernard[6], J. Bock[7], P. De Bernardis[8], J. Charra[2], F. Couchot[9], J. Delabrouille[10], G. Efstathiou[11], M. Giard[6], G. Guyot[2], A. Lange[12], B. Maffei[4], A. Murphy[13], F. Pajot[2], M. Piat[2], I. Ristorcelli[6], D. Santos[14], R. Sudiwala[4], J.F. Sygnet[2], J.P. Torre[2], V. Yurchenko[12], D. Yvon[15],

(1) LERMA, Observatoire de Paris, 61 Bd de l'Observatoire, 75014 Paris
(2) Institut d'Astrophysique Spatiale, Université Paris-Sud, 91405 Orsay Cedex, France
(3) Institut d'Astrophysique de Paris, 98bis Bd Arago, 75014 Paris, France
(4) Cardiff University, Wales, 5 The Parade, Cardiff, UK
(5) CRTBT, CNRS, 53 Av des Martyrs,38026 Grenoble, France
(6) CESR, 9 Av Cl Roche, 31018 Toulouse, France
(7) Jet Propulsion Laboratory, Pasadena, Ca, USA
(8) Universita La Sapienza, Roma, Italy
(9) Laboratoire de l'Accélérateur Linéaire, IN2P3, UPX, 91405 Orsay, France
(10) PCC, Collège de France, Paris, France
(11) Institute of Astronomy, Cambridge, UK
(12) Caltech, Pasadena, Ca, USA
(13) Dept. of Physics, State University of Maynooth, Ireland
(14) IN2P3/ISN, 53 Av des Martyrs,38026 Grenoble, France
(15) DAPNIA/SPP, Commissariat à l'Energie Atomique, Gif sur Yvette, France

**Abstract.** The High Frequency Instrument (HFI) of Planck is the most sensitive CMB experiment ever planned. Statistical fluctuations (photon noise) of the CMB itself will be the major limitation to the sensitivity of the CMB channels. Higher frequency channels will measure galactic foregrounds. Together with the Low Frequency Instrument, this will make a unique tool to measure the full sky and to separate the various components of its spectrum. Measurement of the polarization of these various components will give a new picture of the CMB. In addition, HFI will provide the scientific community with new full sky maps of intensity and polarization at six frequencies, with unprecedented angular resolution and sensitivity. This paper describes the logics that prevailed to define the HFI and the performances expected from this instrument. It details several features of the HFI design that have not been published up to now.

# INTRODUCTION

Cosmic Microwave Background observations have known an amazing and incredibly fruitful development during this last decade. It started with the accurate measurement of the CMB spectrum by the COBE-FIRAS experiment [1] and the first measurement by COBE-DMR [2] of anisotropy other than the dipole. CMB sciences benefit to day from the high quality results of the second generation CMB satellite WMAP [3] and from ground-based high resolution data. These data confirm the pioneering observations of balloon-borne experiments Boomerang[4], Maxima [5] and Archeops [6], and set a consistent base for a standard model in cosmology. In the same time, major issues of cosmology remain open, as stated by several speakers of this conference. Part of the answers is to be found in further observations of the CMB, with improved accuracy and angular resolution.

The main goal of the Planck-HFI instrument is to improve the accuracy of the measurement of the CMB in order to extract cosmological parameters that remain poorly constrained after the results of WMAP and of the best ground-based experiments. The basic idea of HFI is to use all the information contained in the CMB radiation, i.e. to perform a radiometric measurement limited by the quantum fluctuations of the CMB radiation itself. In these conditions, the accuracy is only limited by the number of detectors, by the duration of the observation and by some unavoidable spurious signals made as small as possible. The technology of very low temperature bolometers is opening the opportunity for such a space mission. It was clear from the beginning of the project that the foregrounds, galactic and extragalactic sources, could become the major source of uncertainty at this level of accuracy. It was decided that the same experiment that would produce CMB maps should also be able to measure the foregrounds accurately enough to remove their contribution. The spectral signature of the various components would be used to perform a high quality separation of all the components of the sky emission. This is at the basis of the design of Planck and of the six-bands design of HFI.

An immediate consequence of this choice is that Planck-HFI is also a mission able to measure all the foregrounds in the submillimeter and millimeter ranges, which is of major astrophysical interest. This makes of Planck-HFI both the $3^{rd}$ generation of space CMB experiments and a full sky survey of the IRAS class for longer wavelengths astronomy.

In the first part, we describe the rationale that leads to the specification of the required HFI performances. The second section is focused on a few specific aspects of the HFI design closely related to the scientific objectives. A short view of the expected deliverables concludes this paper.

# DIMENSIONING HFI

## Spectral coverage and number of channels

Consistently with the aim of performing high accuracy measurements of the CMB, the spectral coverage and the number of channels of Planck, and especially of Planck HFI, have been defined to give enough information on the foreground sources, in order to identify them, to measure them, and to subtract their contributions from the map, in a complex component separation process. These foreground sources are formed by the evolving universe situated between us and the warm primordial universe emitting the CMB. Among these, we see the emission of dust and gas in our own galaxy and from other galaxies. Accurate data in the frequency range of Planck will renew the knowledge of dust in our galaxy, and especially of its components at very low temperatures [7]. Clusters of galaxies, that contain high temperature gas detected in the X-rays, distort the CMB by inverse Compton scattering. This is the Sunyaev-Zeldovich Effect (SZE), that makes clusters of galaxies good tracers of the dynamics of the universe at large scales. The measurement of SZE on the cluster A2163 (ref.[8]) is an excellent example of the difficulty to separate the various components that makes the foregrounds. Figure 1 represents the difference of brightness between A2163 and its immediate environment, on a large range of wavelengths.

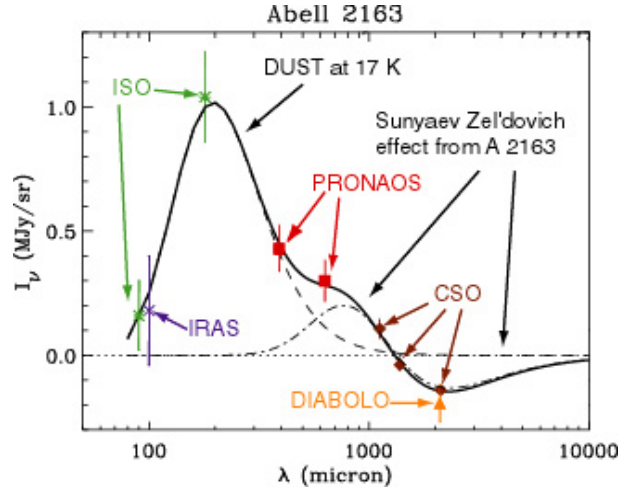

**Figure 1:** Composite spectrum of the excess brightness in the direction of A2163. One can distinguish the negative part of the Sunyaev-Zeldovich Effect ($\lambda>1300\mu m$), its positive part (400 μm<λ<1300 μm), and the contamination by galactic dust (100 μm<λ<1000 μm)

One can distinguish the contribution of the uneven distribution of dust in our galaxy as a positive peak at short wavelengths. This could as well be a negative peak, since it depends on the spatial distribution of dust. that directly contaminates the measurement. This illustrates the need to use a large spectral range to separate the components without ambiguity.

An ambitious modeling of the sky and of the component separation has been undertaken [9] to optimize the design of the instrument, and in particular the number of channels needed for this inversion process. Planck HFI will be a powerful "SZE machine", able to measure SZE on thousands of clusters, which is one of the few tools for the measurement of the large scale properties of the universe.

It was found that six bands in the HFI and three more in the LFI provide enough information to separate the various components by using their spectral and spatial signature. It was shown that, after removal, the residual contamination of the CMB measurement becomes negligible excepted near to the galactic plane, even at the low level set by the sensitivity of HFI.

frontier between the radio receivers of LFI and the bolometers of HFI results from the relative constraints of both technologies. HEMT detectors of LFI are less sensitive at high frequency, due to the quantum limits of any coherent receiver. On the other hand, it becomes more and more difficult to develop bolometers large enough to absorb very long wavelengths. The difficulty to cool large devices at very low temperatures (100mK) was also a major limitation that was taken into account while designing the HFI. The distribution of detectors in frequency is given in table 1.

**Getting sensitivity: Number of detectors per channel**

The basic assumption that lead to the birth of the HFI concept is that submillimeter technology would allow to reach or to approach sensitivities set by the fundamental limits of physics. This defines a new class of experiments, especially for CMB science, since more than two orders of magnitude could be won with respect to COBE-DMR in terms of sensitivity. This enormous ratio could then be employed to increase the number of observed pixels and/or get a better signal to noise per pixel, following optimizations schemes that remain open.

This assumption, made in the 90's, was based on spectacular progresses in all technical fields of submillimeter astronomy with low temperature bolometers that were confirmed in the following years in particular thanks to the research and development activities stimulated by HFI. The initial design of the HFI instrument of Planck was made by assuming that the overall noise of a detection chain would not exceed twice the photon noise of the detected radiation.

This guideline has been kept up to now and has proven to give a realistic approach both with respect to much more sophisticated simulations of the instrument sensitivity and by comparison with experimental results obtained in laboratory or with instruments using similar technologies (Maxima, Archeops, Boomerang). This solid basis was used to develop an end to end simulation of instrument performances, of the measurement of a piece of sky, of the separation of the various components of the foreground and of the inversion of the CMB data. This simulation was used to optimize the instrument and to set the performance requirements. In particular, they have shown that the major CMB scientific objectives could be met even with a sensitivity twice worse than expected, but that a significant part of the objectives, especially extragalactic science, will depend on the achievements of the goal sensitivity.

A few parameters remained to be fixed to be able to compute a real sensitivity on the sky: The efficiency of the optical chain (optical transmission), the thermal emission of the telescope and other optical elements, that could contribute to the total photon noise. The first parameter was fixed at 25%, following the best results demonstrated at this time. We know to-day that it can be as good as 50%. The second parameter (thermal emission of optics) proved to be an essential choice for the project. It was shown[10] that passive cooling of low emissivity telescopes would produce low enough backgrounds in the CMB sensitive region of the spectrum (above 1mm). Temperatures below 60K are required, which could be reached only in orbits far from the earth. This condition was also required for straylight control, the Earth being a major source of straylight at the foreseen level of sensitivity, especially for an experiment on a low earth orbit.

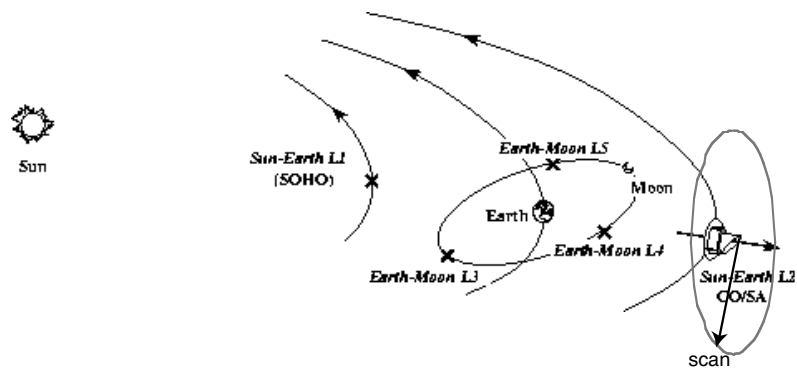

**FIGURE 2.** Schematic representation of the Planck orbit and scanning strategy.

The other major choices were to observe the full sky, to measure polarization, and to measure spherical harmonics up to 3000, which sets beam size to about 5 minutes of arc. A scanning strategy based on large circles on the sky in a plane about perpendicular to the sun, and a mission duration of about one year allowed to define the integration time available for each pixel, and therefore the sensitivity per detector at each observation frequency. It appeared from the beginning that at least 4 detectors per spectral channel were needed to achieve the wanted sensitivity per pixel. This was in good agreement with the need to have some redundancy in each channel.

Measurement of polarization is based on the use of several polarization selective detectors observing the same part of the sky. In principle, only three different orientations of the measured polarization are enough to get the intensity and the Stokes parameters Q and U. Measuring circular polarization (Stokes parameter V) would require a much more complex experimental setup. Since no circular polarization from CMB origin is expected, measuring V was not considered as a goal important enough to justify the induced additional complexity. A pair of orthogonal Polarization Sensitive Bolometers (PSBs) is contained in each PSB mechanical module. The natural configuration is to use four bolometers in two pairs, each pair using a single horn and filter system.

Principles on redundancy and quest for sensitivity lead to the choice of four pairs of PSBs (four horns) per spectral channel (table I). The achieved sensitivities are presented in table 2.

**Table 2:** Expected (goal) mean properties of maps obtained with Planck-HFI

| Central Frequency ($\nu$) | Ghz | 100 | 143 | 217 | 353 | 545 | 857 |
|---|---|---|---|---|---|---|---|
| Spectral resolution | $\nu/\Delta\nu$ | 3 | 3 | 3 | 3 | 3 | 3 |
| Number of unpolarized detectors | | 0 | 4 | 4 | 4 | 4 | 4 |
| Number of polarized detectors | | 8 | 8 | 8 | 8 | 0 | 0 |
| Beam Full Width Half Maximum | arcmin | 9.2 | 7.1 | 5.0 | 5.0 | 5.0 | 5.0 |
| DT/T Sensitivity (Intensity/pixel) | µK/K | 2.8 | 2.2 | 4.8 | 15 | 147 | 6700 |
| DT/T Sens/pixel (U and Q) Polar. | µK/K | 4 | 4.2 | 9.8 | 30 | — | — |
| Total Flux Sensitivity per pixel | mJy | 14.0 | 10.2 | 14.3 | 27 | 43 | 49 |
| ySZ per FOV (x10$^6$) | | 1.8 | 2.1 | 615 | 6.5 | 26 | 605 |

The sensitivity in this table is given in relative thermodynamic temperatures, i.e. as sensitivity to relative changes of CMB temperature. It is given for square pixels with the side equal to the beam FWHM. Sensitivity is also given in Flux units for point sources, and in the dimensionless comptonization factor ySZ for the observation of clusters of galaxies. This table gives the expected sensitivities, which are currently very near to the numbers that had been defined as goals at the beginning of the project. Due to the novel design of HFI and the large uncertainty on the performances of some new subsystems, we have defined requirements based on demonstrated performances, and goals based on expected instrumental progresses. Thanks to the work in the lab and to the demonstration on balloon-borne experiments, it is now considered as most probable that the goal values reported in this table will be met or improved. The main unknown that calls for caution is the possible interaction between the HFI and other systems in the satellite, like telemetry radio emitters, mechanical coolers, or more simply the LFI.

# OPTIMIZING THE DESIGN

## Trade-off between angular resolution and straylight control

Although bolometers are "incoherent" detectors, their coupling with the Planck telescope is achieved through corrugated feed horns, whose performances require a description in terms of mode propagation specific to coherent radio techniques. In the usual way of radio engineering, one has to think to the detector as a source, and to apply the Helmoltz reversion principle to propagate the flux from the receiver to the sky. One can therefore start from the illumination of the telescope produced by the horns that is well approximated by a Gaussian The main parameter that is left for optimization is the width of the illumination of the reflectors by the horns. Increasing this width also increases the illumination of the edge of the mirrors (Edge taper) and the amount of energy that falls outside of the reflectors (spill-over). It also has the effect of improving angular resolution, because the electromagnetic fields in the far field beam and that on the main reflector are basically Fourier pairs. In consequence, to improve angular resolution, one must widen the "illumination" of the main mirror, which also increases the Edge Taper and the "spillover", and therefore the straylight.

This simple physics determines the nature of the trade-off that has to be done between angular resolution and straylight. The acceptable level of straylight is fixed by the requirement that the energy collected from non-CMB sources (mainly the galaxy, the Sun, the Earth) remains less than the instrumental noise. Therefore, the HFI ambitions in terms of sensitivity have the immediate consequence that the requirement on straylight must be tighter than in any previous experiment of the same type.

**Table 3.** Edge taper and spillover in HFI

| Central Frequency | GHz | 100 | 143 | 217 | 353 | 545 | 857 |
|---|---|---|---|---|---|---|---|
| Beam FWHM | Arcmin | 9.2 | 7.1 | 5.0 | 5.0 | 5.0 | 5.0 |
| Edge Taper | dB | -25 | -28 | -32 | -32 | -30 | -30 |
| Spill-over | % | <0.6 | <0.4 | <0.2 | <0.1 | <0.1 | <0.1 |

The design of the horns plays a critical part. While the Gaussian beam pattern is the nearly ideal case, the deviation from Gaussian depends on the details of the horn design. Maffei et al [11] have optimised the HFI horns to obtain the best trade-off between angular resolution and straylight while meeting number of other constraints in terms of mass, relative lengths of horns, diameter, etc… Table 3 gives the levels of the required edge taper and spillover that will be reached thank to the "flared and profiled" design of the HFI horns.

**Focal plane layout: scanning and sampling the sky with individual detectors**

In addition to redundancy, having enough detectors helps getting a correct sampling of the sky in the cross-scan direction. The nominal scanning strategy supposes that every hour, the spin axis of the satellitewill be shifted by 2.5 minutes of arc, which makes the one degree per day needed to approximately follow the Sun-Earth axis. This Delta angle is too large to offer a proper sampling of the sky in channels with a 5 arcmin beam. Choosing a smaller depointing angle would mean increasing the frequency of the depointing, and therefore losing more time for maneuvers and post maneuvers activities. It was decided to stagger the detectors of a same channel by a small amount (1.25arcmin), in order sample the cross scan orientation according to the Nyquist criterion or better.

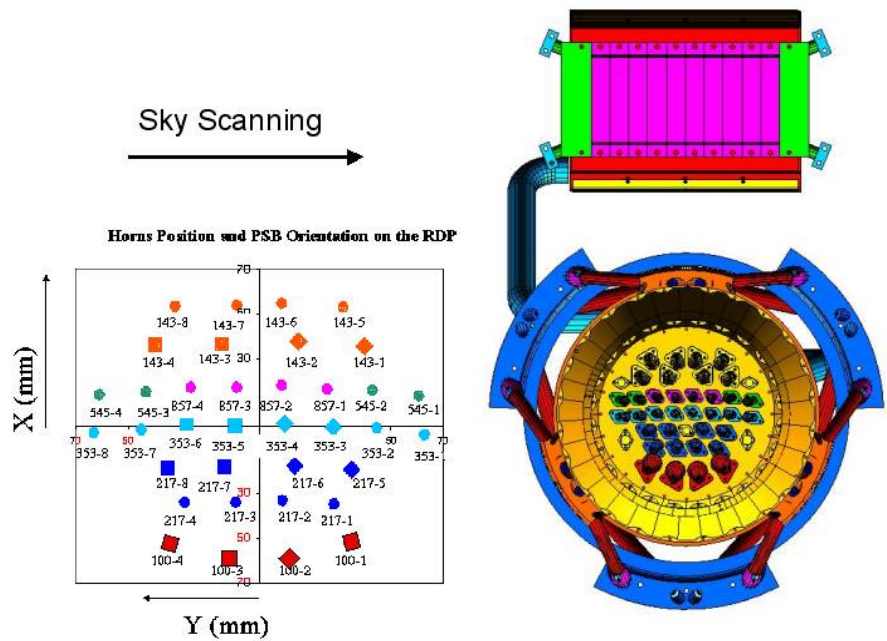

**Figure 3:** Detector layout in the focal plane (left). The focal plane unit (right) houses the detectors. Large horns are for long waves. Note the preamplifier box above the FPU.

The horns that couple the detectors with the sky are distributed following this "staggering" constraint. The high frequency channels are located near to the central part of the focal plane, since they are the most sensitive to aberrations, which are minimal in this place. They are put in a curve that compensates for the distortion of the telescope so that they look at aligned fields of view on the sky, along the scanning orientation.

**Detection chains**

The detectors are bolometers cooled at 100mK constructed by the Jet Propulsion Laboratory in Pasadena. All bolometers use the so called "spider web" principle, which means that the absorption of radiation is achieved thanks to a grid or a mesh of resistive metallic films, while the temperature variation is detected by Doped germanium thermometers. Figure 4 represents three types of bolometers used in HFI. The absorber of the Polarization Sensitive Bolometers are parallel resistive wires, sensitive uniquely to the radiation with the electrical field parallel to the wires. A second bolometer is just behind the first one and detects the orthogonal polarization that is not affected by the first absorber. Therefore, the image of the two PSBs (left bottom picture) looks like a single mesh.

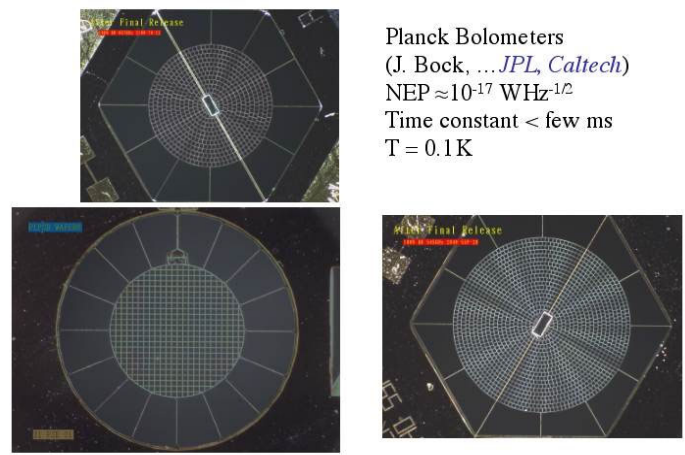

**Figure 4:** Three types of bolometers. The absorber is the mesh at the center of the pictures. Diameters are about twice the wavelength.

**Obtaining a flat noise spectrum**

The scanning strategy consists of performing repetitive circular scans at 1 RPM during one hour and then re-pointing the spin axis of the satellite. The alleviate the problem of de-striping the maps obtained in such a way, the low noise frequencies have been specified to be negligible at time scales shorter than the scan period. It has been shown (Delabrouille) that stripes can be removed without adding significant noise to the maps. All the subsystems have been designed accordingly.

Reaching this kind of very good stability is difficult, since the test conditions and any kind of change in the environment can induce signal fluctuations. The temperature changes of the 100mK base plate of the bolometers can induce signals undistinguishable of real optical signals. A sophisticated thermal architecture has been developed [12] to obtain the 20nKHz$^{-1/2}$ required for consistency with other sources of noise.

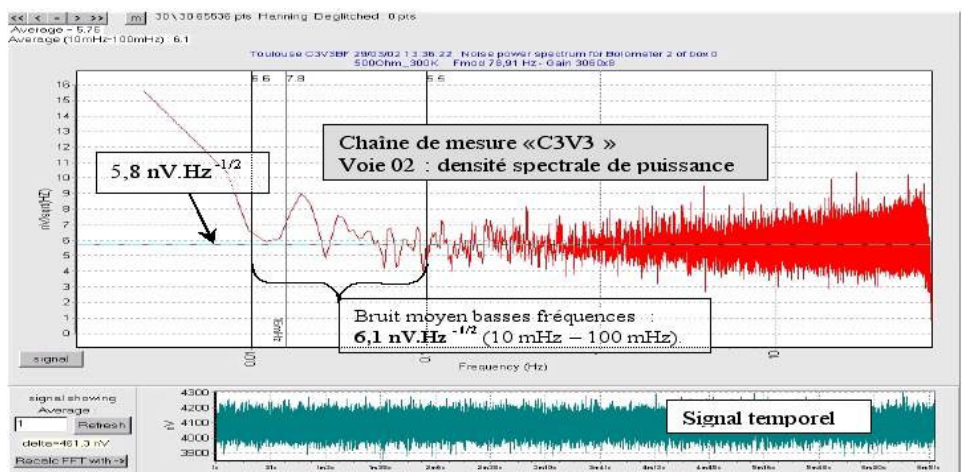

**Figure 5:** Standard presentation of the signal from a readout electronics (lower curve) and of its spectral density (upper curve), showing the low frequency noises.

**DELIVERABLES**

The decomposition in spherical harmonics of the maps expected from Planck let hope a breakthrough in the knowledge of the CMB and of the primordial universe. The accuracy of consistent data obtained with a single experiment will help separate theoretical models that lead to rather similar spectra and that only complete and accurate data will discriminate. Even the B-mode component of the polarization maps will be measurable in reasonably optimistic assumptions.

It must be noted that, due to the scanning strategy the regions near to the ecliptic poles are observed repeatedly during the whole mission. About 1% of the sky is observed with an integration time about 10 times longer than average. The maps of these regions will allow to go deeper in tackling some problems, such as the effect of gravitational lensing, or testing instrumental systematics.

In addition to CMB maps, the Planck mission will produce maps of several components, such as the SZE on thousand of clusters of galaxies, extragalactic sources, dust in the interstellar medium, free-free and synchrotron emission of the galaxy. HFI will also produce maps at six frequencies that are not covered by any other programmed instruments with such accuracy and angular resolution.

**REFERENCES**

[1] J.C. Mather et al. ApJ L, Vol 354, 1990, L37-L40

[2] G.F. Smoot et al., ApJ Letters, vol. 369, 1992, L1-L5

[3] C.L. Bennet et al., ApJ, Vol.583, 2003, pL1-23

[4] P. de Bernardis et al., Nature, Vol.404, 2000, p-p 955-959

[5] S. Hanany et al., ApJ, vol.545, L5-L9

[6] A. Benoit et al., A&A, vol.399, 2003, p.L19-L23

[7] I. Ristorcelli et al., EAS Publications Series, Volume 4, Proc. of IR and Submm Space Astronomy",. EDP Sciences, 2002, pp.9-9

[8] J.M. Lamarre et al., ApJ Vol. 507, 1998, p. L5-L8

[9] Gispert, R.; Bouchet, F. R., Editions Frontieres, 1997. ISBN: 3863322087., p.503

[10] J.M. Lamarre, F.-X. Désert, T. Kirchner, *Background limited infrared and submillimeter instruments*, Space Science Reviews, 74, 1995

[11] B. Maffei et al. Int. J. of IR and MMW, 21, (12) 2023-2033, December 2000

[12] M. Piat et al., IR Space Telescopes and Instruments. Edited by John C. Mather . Proceedings of the SPIE, Volume 4850, pp. 740-748 (2003).